\documentclass[runningheads]{llncs}
\usepackage[T1]{fontenc}
\usepackage{graphicx}
\usepackage{booktabs}
\usepackage{paralist}
\usepackage{subcaption}
\usepackage{wrapfig}
\usepackage{tabularx}
\usepackage{multirow}
\usepackage{appendix}
\usepackage{hyperref}
\usepackage{xurl}
\usepackage{orcidlink}

\usepackage{color}

\begin{document}
\title{Secret Leader Election in Ethereum PoS: An Empirical Security Analysis of Whisk and Homomorphic Sortition under DoS on the Leader and Censorship Attacks}
\titlerunning{Secret Leader Election in Ethereum PoS}
\author{Tereza Burianová \and
Martin Perešíni\orcidlink{0000-0002-2875-9567} \and
Ivan Homoliak\orcidlink{0000-0002-0790-0875}}
\authorrunning{Burianová et al.}
\institute{Brno University of Technology, Faculty of Information Technology, Brno, Czech Republic\\
\email{terezaburianova@seznam.cz}, \email{\{iperesini,homoliak\}@fit.vut.cz}}
\maketitle
\begin{abstract}
Proposer anonymity in Proof-of-Stake (PoS) blockchains is a critical concern due to the risk of targeted attacks such as malicious denial-of-service (DoS) and censorship attacks. 
While several Secret Single Leader Election (SSLE) mechanisms have been proposed to address these threats, their practical impact and trade-offs remain insufficiently explored. 
In this work, we present a unified experimental framework for evaluating SSLE mechanisms under adversarial conditions, grounded in a simplified yet representative model of Ethereum’s PoS consensus layer. 
The framework includes configurable adversaries capable of launching targeted DoS and censorship attacks, including coordinated strategies that simultaneously compromise groups of validators. 
We simulate and compare key protection mechanisms -- Whisk, and homomorphic sortition.
To the best of our knowledge, this is the first comparative study to examine adversarial DoS scenarios involving multiple attackers under diverse protection mechanisms.
Our results show that while both designs offer strong protection against targeted DoS attacks on the leader, neither defends effectively against coordinated attacks on validator groups. 
Moreover, Whisk simplifies a DoS attack by narrowing the target set from all validators to a smaller list of known candidates.
Homomorphic sortition, despite its theoretical strength, remains impractical due to the complexity of cryptographic operations over large validator sets. 

\end{abstract}
\section{Introduction}
The transition of Ethereum from a Proof-of-Work (PoW)~\cite{nakamoto} to a PoS~\cite{ethPoS} consensus mechanism marked a significant milestone in the evolution of blockchain technology.
Unlike PoW systems, where computational effort is used to determine block proposers, PoS selects proposers based on staked value, aiming to reduce energy consumption and increase scalability.
However, this architectural shift introduces new attack surfaces, particularly concerning the predictability and exposure of block proposers~\cite{packetology,HOPRsniping,algorand}.

In Ethereum’s PoS, validators are selected in advance to propose blocks using the RANDAO~\cite{randao} randomness beacon.
Although this design enables validators to prepare for their duties, it inadvertently exposes their schedule of block proposals to adversaries.
Since proposer identities are publicly known prior to their assigned slots, attackers can target them with Denial-of-Service (DoS) attacks or apply censorship tactics~\cite{DoSMagicians,CensorshipTornadoAnalysis}.
These attacks can degrade network performance, cause economic loss, or even lead to validator centralization by selectively penalizing certain participants~\cite{DoSMagicians}.
By targeting specific proposers who might include undesirable transactions (e.g., OFAC non-compliant relays in 2022~\cite{mevwatch}), adversaries can enforce censorship, weaken liveness, or economically coerce validators, threatening the integrity of the network~\cite{CensorshipTornadoAnalysis}. 
The visibility of proposers is particularly problematic in scenarios involving Maximum Extractable Value (MEV)~\cite{mev}, as it allows attackers to target block proposers from previous slots, expanding their MEV opportunities and gaining financial advantages~\cite{FlashbotsDrake}.

A variety of mechanisms have been proposed to protect proposers in PoS-based systems, ranging from SSLE mechanisms~\cite{SSLEBoneh} like Whisk~\cite{Whisk} and Blind-and-Swap~\cite{swapornot}, to homomorphic sortition~\cite{hSortition}, VRF-based protocols such as Algorand~\cite{algorand}, and decentralization through Distributed Validator Tech\-nolo\-gy (DVT) \cite{dvt}. 
While approaches like DVT, network-layer solutions~\cite{dandelion}, and SnSLE~\cite{SnSLE,algorand} face drawbacks related to complexity, latency, or compatibility, SSLE mechanisms offer protocol-level proposer anonymity with minimal assumptions, making it the most promising fit for Ethereum’s consensus. 
Complementary mechanisms like Proposer-Builder Separation (PBS)~\cite{pbs,epbs} further mitigate incentives for targeting proposers, though they do not provide strict anonymity.

This paper focuses on two SSLE mechanisms: (1) Whisk~\cite{Whisk}, a shuffle-based SSLE protocol leveraging zero-knowledge proofs (ZKPs) and (2) homomorphic sortition~\cite{hSortition}, which leverages fully homomorphic encryption (FHE) for collaborative encrypted leader election.
While conceptually different, both approaches share the goal of mitigating adversarial visibility of proposers.

To evaluate these mechanisms, we developed  a simulation framework that models the consensus behavior of Ethereum~\cite{consensusSpec} and integrates Whisk and homomorphic sortition.
Several adversarial scenarios were designed to test the effectiveness of these protections under conditions such as DoS attacks and targeted censorship.
The performance of each mechanism was assessed not only in terms of efficacy but also with respect to computational overhead and feasibility within the strict 12-second slot timing of Ethereum.

\paragraph{\textbf{Contributions}}
The contributions of this paper are as follows:
\begin{compactitem}
    \item[$\bullet$] We made a threat analysis of Ethereum's PoS proposal selection, identifying specific vectors such as block proposer censorship and DoS.
    \item[$\bullet$] We made a comparative study and simulation-based evaluation of two proposer protection mechanisms (Whisk and homomorphic sortition).
    \item[$\bullet$] We elaborated on results, trade-offs, future directions, including integration into the existing Ethereum protocol, and open challenges in scalability.
\end{compactitem}

\paragraph{\textbf{Organization}}
The paper is structured as follows. \autoref{background} provides background on blockchain architecture, consensus mechanisms, and Ethereum’s PoS.
\autoref{sec:threat-model} describes attacks we dealt with.
\autoref{mechanisms} describes the chosen protection mechanisms, focusing on their cryptographic foundations and operational logic.
\autoref{framework} outlines the simulation framework and the attack scenarios used for evaluation.
\autoref{analysis} presents the results of our experiments. 
\autoref{sec:future-work} discusses the results, their implications, and outlines directions for future research. 
\autoref{relatedwork} provides an overview of related work and \autoref{conclusion} concludes the paper.

\section{Background on Ethereum}\label{background}
Ethereum is a general-purpose, permissionless blockchain platform that supports decentralized applications and smart contracts~\cite{yellowpaperEth}.
It operates on a layered architecture composed of a network/data layer, consensus layer, execution layer and application layer.
Communication between participants occurs over a peer-to-peer network, while consensus mechanisms ensure agreement on the global state of the chain.

\paragraph{\textbf{Proof-of-Stake in Ethereum.}}
In Ethereum’s PoS consensus, validator identities are represented by a staked amount of Ether (ETH).
Validators participate by proposing blocks, attesting to others' proposals, and finalizing the chain state.
Instead of competing for block production using computational power (as in PoW), validators are deterministically selected to perform these duties based on a shared source of randomness and their staked value~\cite{ethPoS}.

\paragraph{\textbf{RANDAO and Proposer Selection.}}
Proposer selection in Ethereum relies on RANDAO, a collaborative randomness beacon.
With each block, proposers contribute unpredictable and verifiable values.
For each epoch, these values are aggregated into a RANDAO mix and stored in the beacon state~\cite{randao}. The randomness seed used for proposer selection is derived from the RANDAO mix two epochs prior, allowing validators to know in advance whether they are selected for the role.
Since the proposer selection is deterministic and public, anyone can predict the proposer for any upcoming slot once the relevant RANDAO mix is available~\cite{annotSpec}.
Such a transparency introduces a vulnerability: adversaries can target known proposers.
Note that we will refer to this version of leader election as the status quo throughout the paper.

\begin{figure}[t]
  \centering
  \includegraphics[width=0.8\linewidth]{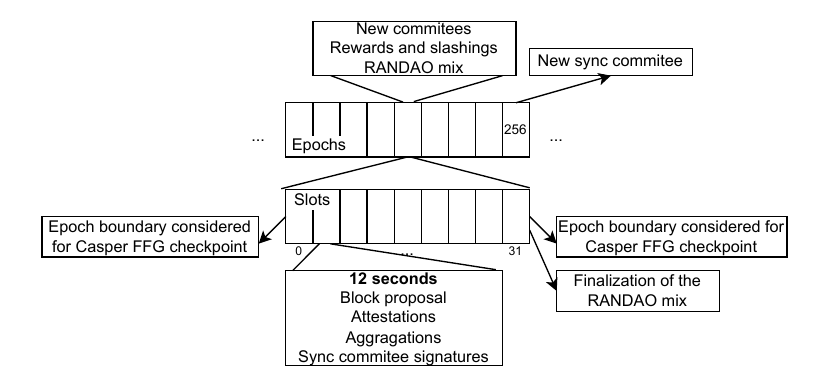}
  \caption{An overview of the Ethereum consensus timeline.}
  \vspace{-0.5cm}
  \label{fig:timeline}
\end{figure}

\paragraph{\textbf{Slot and Epoch Timeline}.}
Ethereum divides time into fixed intervals: slots (12 seconds) and epochs (32 slots).
Each slot ideally includes a new block proposal by a designated validator and attestation duties by a validator subset.
Validators are informed of roles two epochs in advance.
Epoch-bound events include RANDAO mix updates, new committees, rewards, slashings, and Casper FFG~\cite{casper} justification/finaliza\-tion checks.
This rigid timeline, shown in \autoref{fig:timeline}, supports fast block finality and predictable scheduling.

\section{Threat Model}\label{sec:threat-model}
Since proposer identities in Ethereum are deterministically derived and announced ahead of time, attackers can tailor targeted actions towards specific validators.
The most relevant threats include:

\begin{compactitem}
\item[$\bullet$]{\textbf{Denial-of-Service (DoS) Attacks.}} An attacker disrupts a validator’s ability to propose a block by flooding its node with traffic.
This can be highly effective if the attacker can identify the proposer in advance and link their identity to an IP address~\cite{DoSMagicians}. Some attacks do not require IP mapping but instead exploit known bugs or vulnerabilities in execution clients (e.g., Geth), allowing disruption of proposers at scale~\cite{GethVuln,DogeReaper}.
DoS attacks in Ethereum can take different forms depending on their target and strategy:

\begin{enumerate}
    \item A \textbf{large-scale DoS attack} targeting the Ethereum network as a whole, rather than individual proposers, potentially motivated by competing platforms seeking to undermine Ethereum, regulatory bodies attempting to restrict its operation, or actors aiming to manipulate cryptocurrency markets by eroding trust through network disruptions.
    \item A \textbf{targeted DoS attack} aimed at validators based on their assigned slots. A notable example occurs with MEV opportunities: by attacking validators in specific slots, such as those immediately before the attacker’s own, an adversary can extend the MEV available to them while reducing fairness and decentralization~\cite{FlashbotsDrake}.
\end{enumerate}

\item[$\bullet$]{\textbf{Censorship.}} Adversaries can prevent specific proposers from including certain transactions, particularly those involving sanctioned addresses~\cite{CensorshipTornadoAnalysis}.

\end{compactitem}
These threats motivate stronger proposer protection mechanisms that preserve anonymity and limit such attacks~\cite{FlashbotsDrake}.

\section{Investigated Approaches}\label{mechanisms}
Several mechanisms for proposer protection in PoS protocols have been proposed, including VRF-based sortition as in Algorand~\cite{algorand}, network-layer obfuscation such as Dandelion$++$ with RLN~\cite{dandelion}, and various SSLE schemes~\cite{SSLEBoneh}. 
Algorand hides proposer identities but selects multiple leaders per round, adding coordination overhead, while network-layer solutions require fundamental changes to Ethereum’s stack. 
In contrast, SSLE mechanisms conceal the leader’s identity until proposal time, unlike Ethereum’s status quo of publicly revealing a leader in advance (see \autoref{background}), and does so without major overhead or incompatibility. 
SSLE mechanisms differ in approach from shuffle-based anonymization to encrypted computation, and in this work we focus on two promising instances: Whisk~\cite{Whisk} and homomorphic sortition~\cite{hSortition}.

\subsection{Whisk: Shuffle-Based Secret Leader Election}
\begin{figure*}[t]
  \centering
\includegraphics[width=\linewidth]{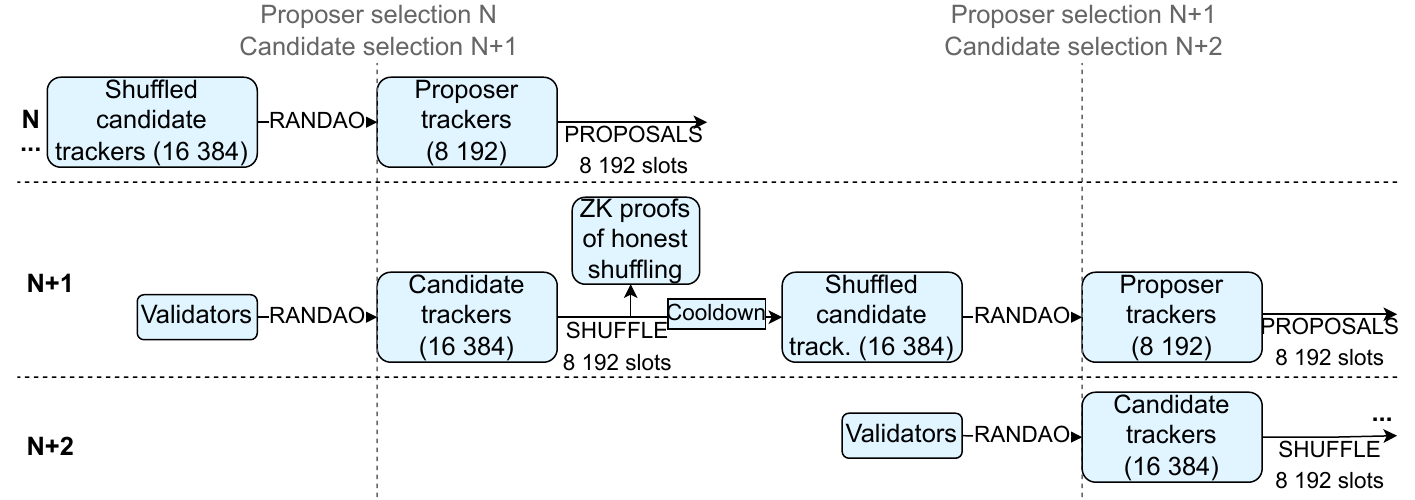}
  \vspace{-0.4cm}
  \caption{The pipeline of the Whisk proposer protection mechanism.}
  \label{fig:whisk}  
  \vspace{-0.3cm}
\end{figure*}

Whisk is an SSLE mechanism designed for Ethereum’s PoS~\cite{Whisk}. It conceals block proposers’ identities until their assigned slot by shuffling a candidate list using verifiable random permutations and ZKPs. Operating in a pipelined mode (shown in~\autoref{fig:whisk}), current-round proposers prepare the shuffle for the next round, enabling scalability while preserving pre-slot anonymity.

\paragraph{\textbf{Commitment and Anonymity.}}
In Whisk, validators commit to a long-term secret $k$ using a tracker $(rG,krG)$, which can be re-randomized by third parties as $(zrG,zkrG)$ without linking to the original. The proposer proves ownership of the tracker using $k$ and ZKPs, handled via Curdleproofs~\cite{CurdleproofsDoc}.

\paragraph{\textbf{Candidate Selection and Shuffling.}}
The process starts by selecting 16,384 candidates from the active validator set using RANDAO, whose identities are initially linkable since trackers are not yet re-randomized. Over 8,192 slots, proposers iteratively shuffle and re-randomize 128 candidate trackers per iteration, with ZKPs via Curdleproofs~\cite{CurdleproofsDoc} ensuring correctness.

\paragraph{\textbf{Proposer Selection.}}
After the iterative shuffling is complete, a second round of RANDAO-based selection is used to choose 8,192 proposers from the shuffled and randomized candidate list, mapping them to the next 8,192 slots.
Concurrently, a new candidate selection and shuffling phase is initiated to prepare the proposer set for the following round, ensuring a continuous pipeline of securely randomized leader selection~(\autoref{fig:whisk}).

\subsection{Homomorphic Sortition: Encrypted Collaborative Selection}
Homomorphic sortition is a cryptographic mechanism that leverages threshold fully homomorphic encryption (ThFHE) to perform proposer selection over encrypted data~\cite{hSortition}. 
This approach ensures that proposers cannot be identified until a joint decryption is completed.

\begin{figure}[t]
    \centering
    \begin{subfigure}{0.45\columnwidth}
        \centering
        \includegraphics[width=\columnwidth]{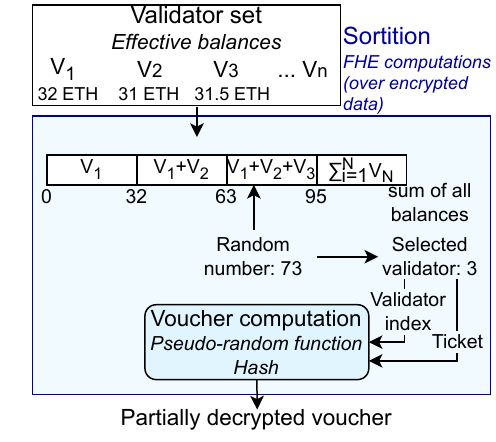}
        \caption{The sortition operation.}
        \label{fig:HSsortition}
    \end{subfigure}
    \begin{subfigure}{0.45\columnwidth}
        \centering
        \includegraphics[width=\columnwidth]{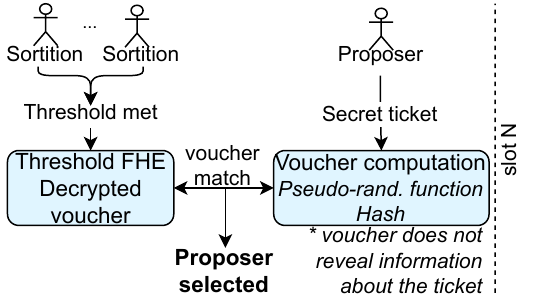}
        \caption{Proposer selection for a single slot.} 
        \label{fig:HSselection}
    \end{subfigure}
    \caption{The processes used in the homomorphic sortition protocol.}
    \vspace{-0.5cm}
\end{figure}

\paragraph{\textbf{Sortition.}} In homomorphic sortition, leader election begins with sortition (shown in \autoref{fig:HSsortition}), where a prefix sum over participants’ stakes forms weighted intervals. An encrypted random number is compared against this encrypted array, and the leader is selected based on the interval in which the number falls. All computations use fully homomorphic encryption (FHE) to preserve privacy. An encrypted voucher is then generated from the proposer’s encrypted ticket.

\paragraph{\textbf{Circuits and Parallel Computation.}}
Homomorphic sortition relies on a set of encrypted circuits capable of performing basic computations--such as addition, comparisons, and hashing directly over ciphertexts. 
These circuits are optimized for parallel execution.

\paragraph{\textbf{Verifiability and Voucher Matching.}}
Once the voucher is decrypted by a threshold of participants and published, any validator claiming to be the elected leader can present a proof of ownership by locally computing the voucher using their secret ticket and demonstrating that it matches the revealed voucher without revealing any secret values, as shown in \autoref{fig:HSselection}.

\section{Simulation Framework \& Implementation}\label{framework}
To assess the viability of proposer protection mechanisms in Ethe\-reum’s PoS consensus, both Whisk and homomorphic sortition were implemented in a custom-built simulation framework. 
The simulation was designed to emulate Ethereum's block proposer workflow under various adversarial conditions, enabling both functional validation and security evaluation under attack scenarios.

\subsection{Simulation Framework Design}
The simulation framework closely mirrors the Ethereum consensus as described in its specifications~\cite{consensusSpec} and partially in \autoref{background}. 
Importantly, it is designed to be modular and extensible, enabling integration of additional proposer anonymity mechanisms for future evaluation.

\paragraph{\textbf{Simulation Parameters.}} The number of validators and the duration of the simulation are configurable, allowing for flexible experimentation. 
The framework supports running validators under both the default Ethereum proposer selection mechanism and the two implemented protection schemes. 
Furthermore, the framework models adversarial behavior through configurable attack modules capable of launching denial-of-service and censorship attacks, allowing for controlled testing of each mechanism's resilience.
The attack configuration includes parameters that define how likely an attacker is to correctly link a validator’s identity to their IP address and how well each validator is protected against attacks. 
By adjusting these, the model can represent different attacker strategies -- such as those relying on standard DoS methods with uncertain targeting, more sophisticated attackers exploiting client vulnerabilities~\cite{GethVuln,DogeReaper}, or adversaries with varying levels of resources and precision.

\paragraph{\textbf{Simulation Model Design.}} The simulation framework was designed to be as simple as possible while still retaining the functionality required for a meaningful analysis.
The final model captures the essential elements of Ethereum’s consensus layer: honest validators proposing blocks according to the specification, slot and epoch processing (including missed slots), proposer selection, basic block proposal processing and RANDAO.

For clarity and focus, we abstract away components of Ethereum consensus and networking that are not relevant to our objectives. 
In particular, this includes details of the communication model, such as bandwidth limitations, latencies, and message propagation. 
For the same reason, we exclude finality, fork choice, transactions and their interaction with the execution layer, balance changes (rewards and slashing), deposits and exits. By abstracting from the network layer, we obtain a synchronous consensus model that isolates the effect of proposer protection mechanisms at the consensus level. Detailed images illustrating the framework’s design and implementation are shown in \autoref{appendix:implementation}.

\subsection{Simulation Framework Implementation}
We implemented a simulation framework in Rust, leveraging its performance and rich ecosystem of Ethereum-related libraries. 
Core beacon structures were abstracted using generic traits to support multiple proposer selection mechanisms and attacks within a unified simulation environment. 
The implementation includes support for cryptographic primitives such as BLS signatures, with serialization handled via \texttt{ssz\_rs} to preserve the Ethereum consensus structure compatibility. 
The framework accepts configuration via JSON, allowing flexible specification of simulation parameters and outputs detailed results for analysis.
We provide the source code for our implementation at \url{https://anonymous.4open.science/r/leader-election-sim-anon-FCFD/}.

\paragraph{\textbf{Implementation of Whisk.}}
The Whisk protocol was implemented with its full pipeline design, where each round of proposers performs the shuffling for the next round of proposer selection, based on the Whisk EIP-7441 specification draft~\cite{eip7441}. 
The number of validators selected for each phase was proportionally reduced to handle reduced validator set size.

Validators were represented using trackers which could be re-randomized, producing unlinkable identities. 
These could later be claimed using a proof that connects the randomized tracker to their original long-term commitment. 
The simulation included ZKPs to verify honest shuffles. 
This was achieved by incorporating an existing Curdleproofs implementation~\cite{curdleproofs} into the pipeline.
The pipelined structure was successfully tested, ensuring that the protocol could maintain continuous proposer assignment while preserving pre-slot anonymity.

\paragraph{\textbf{Implementation of Homomorphic Sortition.}}
Homomorphic sortition was implemented using simplified Fully Homomorphic Encryption (FHE) constructs, eliminating the threshold functionality to simulate the behavior of a single proposer. 
Each validator was represented by a ticket, a random encrypted number. A list with intervals based on validators' effective balances was created, and a voucher, representing the proof of the selected proposer, was computed based on the random selection from the list. 
The proposer then computes the voucher without encrypted operations, comparing it to the now decrypted voucher computed in the sortition process.

The implementation focused on maintaining the structure and computational steps of the protocol. 
To implement the FHE circuits, TFHE-rs by Zama~\cite{TFHE-rs} was used. To implement the pseudo-random function and the hash, a simplified Prince block cipher~\cite{prince} was implemented. 
A hash function based on Hirose's construction~\cite{hirose}, utilizing two instances of the PRINCE block cipher, was used to compute the hash.
There are two options to launch the homomorphic sortition simulation. 
The simplified version omits the voucher verification and uses the pseudorandomness provided by the TFHE-rs implementation for longer simulations. 
The full version utilizes the implemented cipher and performs all checks. This version serves as a frame for functionality checks and cost measurements.

\paragraph{\textbf{Model Validation.}}
We validated the functionality of the implemented models by running baseline experiments (i.e., without attacks). 
These confirmed that proposers were correctly assigned to their slots, the selection process remained fair, and validator identifiers were linkable to their identities under the status quo. In contrast, under the implemented protection mechanisms, the identifiers remained secure and unlinkable to the validators’ real identities.

\subsection{Simulated Attack Types}
This section describes how the threats detailed in \autoref{sec:threat-model} are incorporated into the framework implementation.

\paragraph{\textbf{DoS Attack.}}
Our simulation framework implements several potential attack vectors targeting Ethereum proposers. 
The implemented \textbf{targeted Denial-of-Service} (DoS) attack aims to disrupt block production by targeting validators based on identified IP-public key pairs, with success rates influenced by correct pairing probability \cite{HOPRsniping} and individual node protection measures like VPNs, firewalls or the more advanced \texttt{sentry} nodes \cite{HardeningBlockchain,DoSMagicians}. 
An \textbf{advanced DoS} variant circumvents the proposer identification by indiscriminately attacking a subset of validators, thereby enabling scalable disruption. 

\paragraph{\textbf{Censorship Attack.}}
In the censorship attack, we selectively target proposers belonging to a victim set, performing DoS only when a victim is scheduled to propose a block. 
This strategy also includes an \textbf{advanced censorship} variant, where the attackers continuously DoS validators from the victim group, irrespective of their immediate proposer status. 
These configurable attack modes enable the precise simulation of targeted disruptions and support the assessment of proposer protection mechanisms under diverse threat scenarios.

\section{Evaluation}\label{analysis}
Experiments were conducted to model a DoS scenario and a censorship scenario in which the attacker was able to correctly associate 80~\% of validators with their IP addresses, while 20~\% of validators had additional protection mechanisms in place against such attacks. 
The simulation was run with a reduced validator set~--~1,000 validators for Whisk and 100 validators for homomorphic sortition. 
We simulated a resourceful attacker capable of targeting 10~\% of the validator set using an advanced DoS strategy. 
In the censorship scenario, 10~\% of the validators were designated as censorship targets. 
The duration of the simulation was 6~epochs with several randomized runs for more statistically significant results.

\subsection{Security Effectiveness}

 \begin{figure}[t]
        \centering        
        \begin{subfigure}{0.45\columnwidth}
            \centering
            \includegraphics[width=\columnwidth]{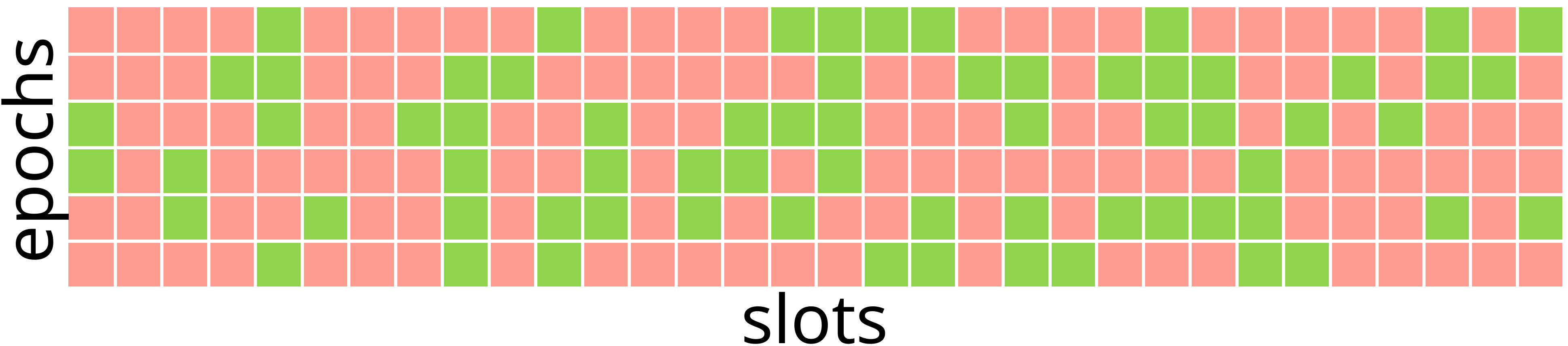}
            \caption{Targeted DoS (no protection).}
        \end{subfigure}
        \begin{subfigure}{0.45\columnwidth}
            \centering
            \includegraphics[width=\columnwidth]{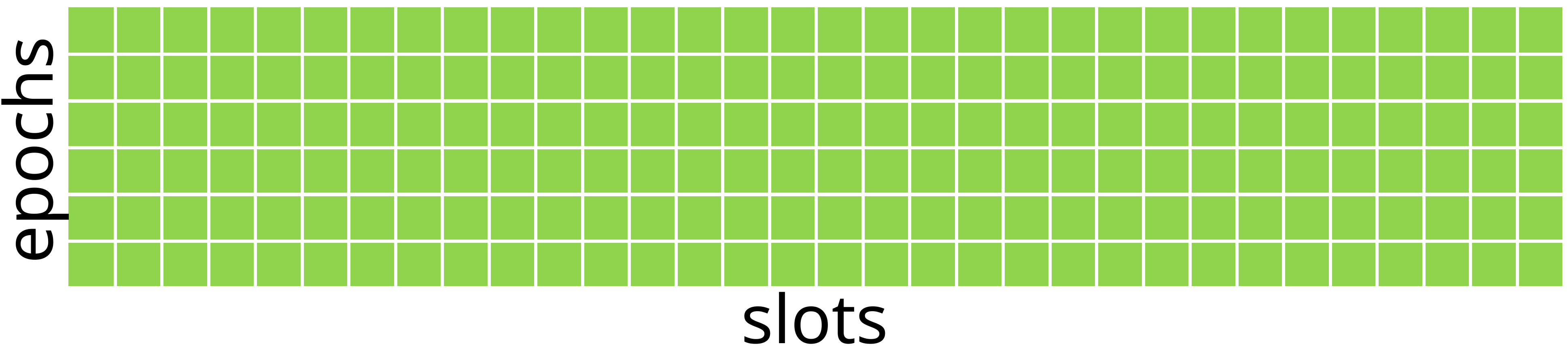}
            \caption{Targeted DoS (with protection).}
        \end{subfigure}
        \begin{subfigure}{0.45\columnwidth}
            \centering
            \includegraphics[width=\columnwidth]{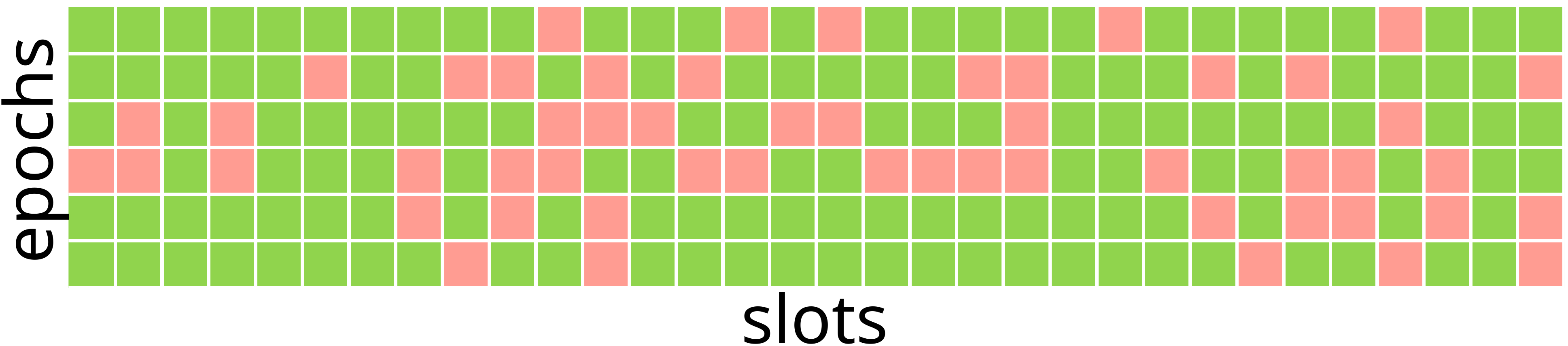}
            \caption{Advanced DoS using Whisk.}
        \end{subfigure}
        \begin{subfigure}{0.45\columnwidth}
            \centering
            \includegraphics[width=\columnwidth]{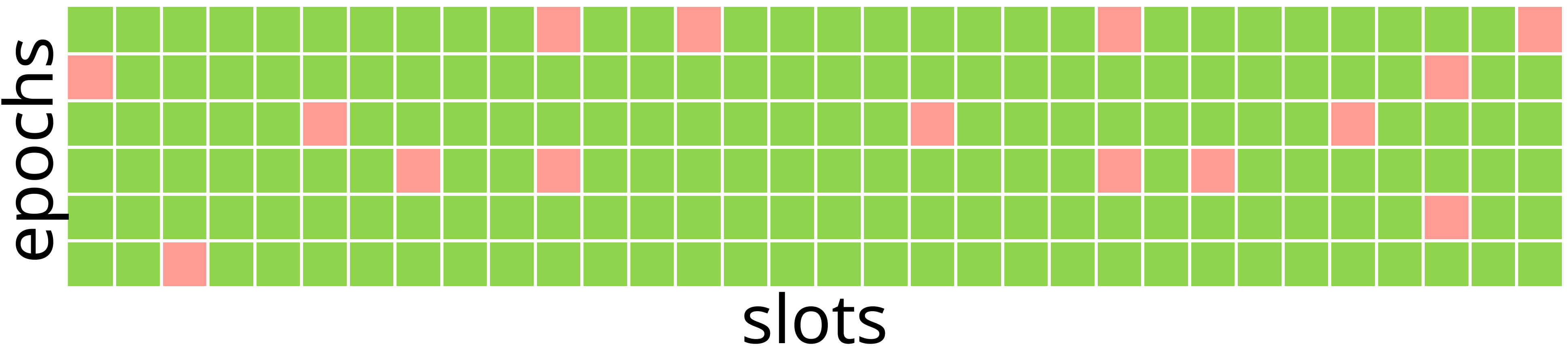}
            \caption{Advanced DoS using h. sortition.}
        \end{subfigure}
        \caption{Comparison of beacon chain activity under different proposer protection mechanisms and attack scenarios, where {\color{green}{green}} indicates successfully proposed blocks and {\color{red}red} indicates missed slots.}
        \label{fig:chainActivity}
    \end{figure}
    
\paragraph{\textbf{DoS Attacks on Leader.}}
The results of the simulation experiments indicate that while current proposer protection mechanisms, such as Whisk and homomorphic sortition, offer promising improvements over the default Ethereum selection, they are not universally robust against all forms of attack.
\autoref{fig:chainActivity} presents data from a single simulation instance, serving as an illustrative example of possible effects. The figure shows that the targeted DoS attack severely impacts block proposal rates when no protection is in place, with 64~\% of blocks missed and 67~\% of proposers affected. 
Both Whisk and homomorphic sortition mitigate targeted attacks by obscuring the identity of the proposer until slot execution, reducing the impact of the targeted DoS to a negligible level close to 0~\%. 
However, advanced DoS attacks remain problematic, particularly for Whisk, which introduces a vulnerability in its candidate selection phase. 
Because the candidate set is known in advance and the candidates' identities are not concealed at this stage, an attacker with sufficient resources can target a large number of candidates, leading to partial or complete disruption. 
This makes the advanced DoS attack potentially more damaging for Whisk,  resulting in 28~\% of blocks missed, compared to 6–8~\% under no protection or homomorphic sortition, where the attacker selects victims from the entire validator set.

\begin{figure}[t]
    \centering
    \begin{subfigure}{0.45\columnwidth}
        \centering
        \includegraphics[width=\columnwidth]{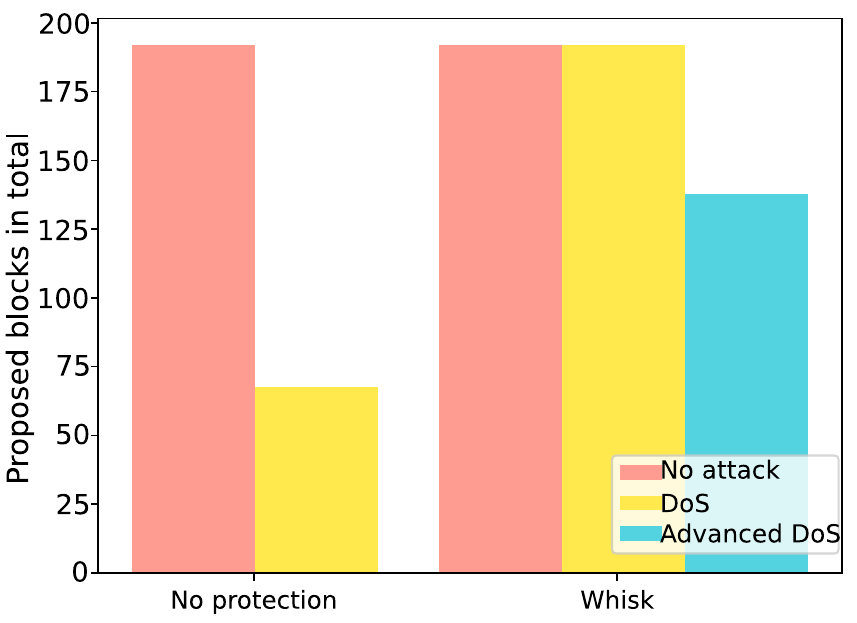}
        \caption{Whisk.}
    \end{subfigure}
    \begin{subfigure}{0.45\columnwidth}
        \centering
        \includegraphics[width=\columnwidth]{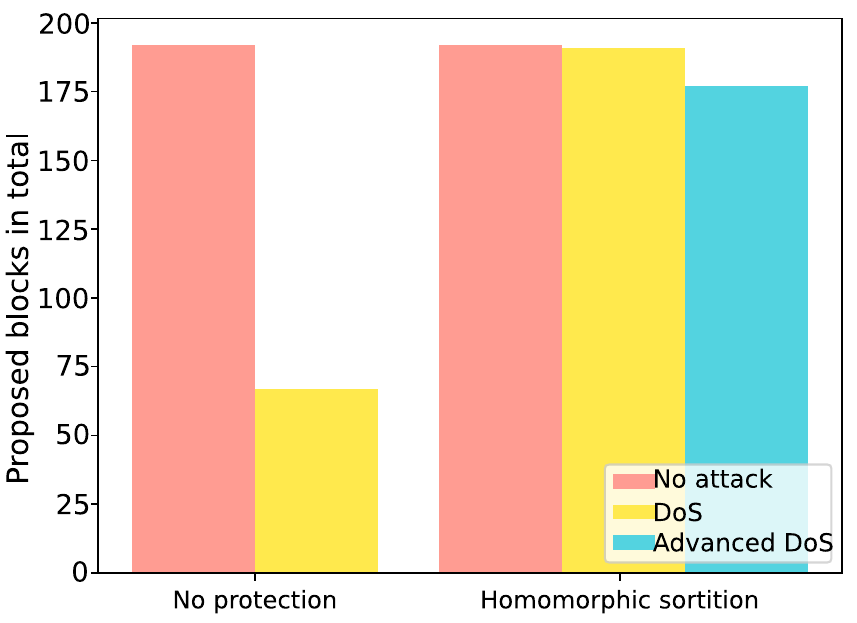}
        \caption{Homomorphic sortition.}
    \end{subfigure}
    \caption{Comparison of the number of proposed blocks using different selection mechanisms under various circumstances during a DoS attack.}
    \vspace{-0.5cm}
    \label{fig:DoSbars}
\end{figure}

\autoref{fig:DoSbars} highlights the difference in the number of successfully proposed blocks under different protection mechanisms and attack scenarios, providing a statistical view of the system’s behavior.
For the "No protection" and "Whisk" configurations, results are aggregated over five simulation runs with different random seeds. Due to the significantly higher computational cost of homomorphic sortition, that configuration was simulated only once. 
Whisk significantly reduces the success rate of standard DoS attacks, but its effectiveness declines under advanced adversarial conditions.
Homomorphic sortition performs similarly to the status quo during advanced DoS since the computations are performed on the whole set of validators, indicating that while it introduces no new vulnerabilities, it also does not improve resilience. 
In the case of targeted DoS, the attack was only effective when the adversary randomly guessed the proposer and successfully eliminated them -- a strategy with negligible impact at realistic validator scales, such as one million participants.

\begin{figure}[t]
        \centering
        \begin{subfigure}{0.45\columnwidth}
            \centering
            \includegraphics[width=\columnwidth]{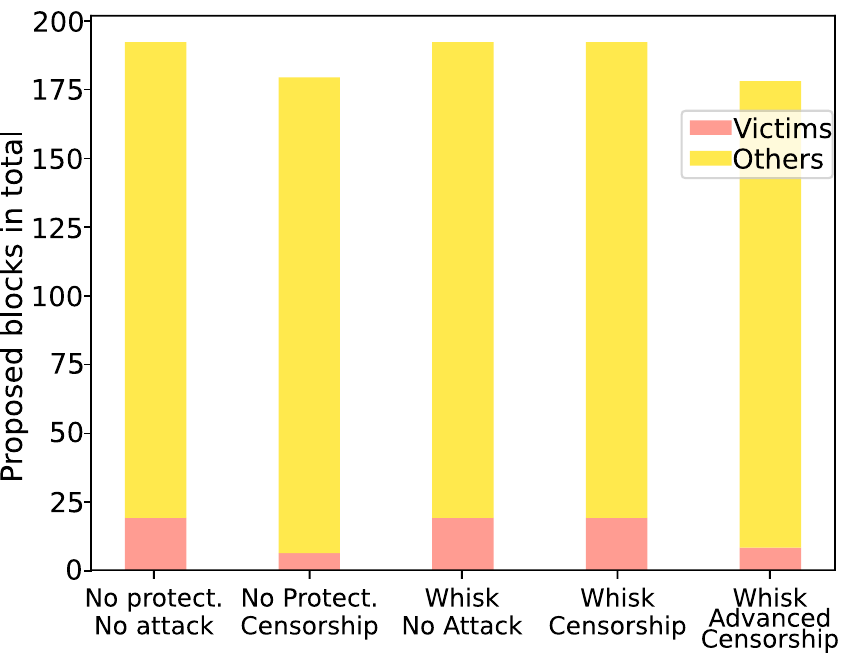}
            \caption{Whisk.}
        \end{subfigure}
        \begin{subfigure}{0.45\columnwidth}
            \centering
            \includegraphics[width=\columnwidth]{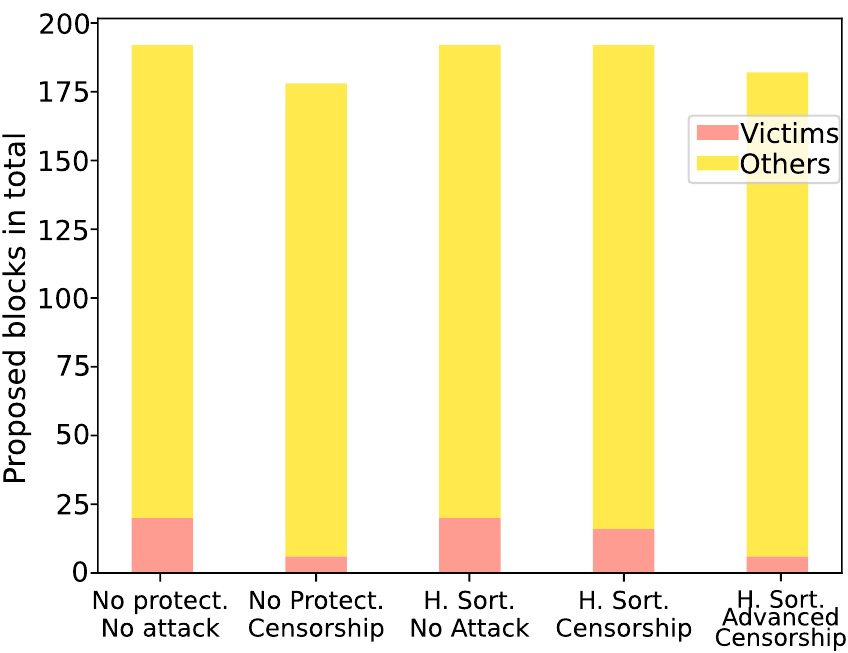}
            \caption{Homomorphic sortition.}
        \end{subfigure}
        \caption{Comparison of the number of proposed blocks using different selection mechanisms under various circumstances during a censorship attack.}
        \vspace{-0.5cm}
        \label{fig:Censorbars}
    \end{figure}

\paragraph{\textbf{Censorship Attack.}}
\autoref{fig:Censorbars} provides a statistical overview of the outcomes, comparable to the structure of the DoS attack analysis in \autoref{fig:DoSbars}. However, in contrast to DoS, where the impact is readily observable through increased missed slots, censorship manifests more subtly and requires deeper analysis of slot behavior and victim targeting.

In terms of observable proposal activity, the difference between scenarios appears negligible. Only around 6~\% to 7~\% of slots are missed during the censorship attack, a deviation that could be attributed to typical validator inactivity, network delays, or benign causes. This makes censorship nearly indistinguishable from regular network fluctuations when viewed at a system-wide level.

However, when examining the effect on the censorship victims set, a stark contrast emerges: approx. 69~\% to 71~\% of targeted validators are prevented from proposing their blocks. This highlights the effectiveness of the censorship strategy, although its footprint on overall slot availability remains minimal.

Both Whisk and homomorphic sortition demonstrated strong protection\break against the targeted censorship strategy by concealing proposer identities until the slot execution. As a result, targeted censorship was rendered largely ineffective under these mechanisms. In contrast, under the advanced censorship strategy, where the attacker persistently targets the entire victim group regardless of their current proposer status, the protective effect diminishes and the outcomes remain close to the no-protection baseline, indicating that the protection is insufficient.

\begin{table}[t]
\scriptsize
\centering
\caption{Time measurements of various consensus processing phases.}
\label{tab:cost}
\vspace{-0.2cm}
\begin{tabular}{ccrrrr}
\\\toprule
\textbf{} & \multicolumn{1}{l}{} & \multicolumn{3}{c}{\textbf{\begin{tabular}[c]{@{}c@{}}Block Processing  {[}ms{]}\end{tabular}}} & \multicolumn{1}{c}{\textbf{\begin{tabular}[c]{@{}c@{}}Epoch\\ processing\\{[}ms{]}\end{tabular}}} \\
\cmidrule(lr){3-5} 
\cmidrule(lr){6-6}
\multicolumn{1}{l}{\textbf{Mechanisms}} & \multicolumn{1}{l}{\textbf{}} & \multicolumn{1}{c}{\textbf{10 val.}} & \multicolumn{1}{c}{\textbf{100 val.}} & \multicolumn{1}{c}{\textbf{1 000 val.}} & \multicolumn{1}{c}{\textbf{-}} \\
\midrule
\multirow{2}{*}{\textbf{Base}} & avg. & 8 & 10 & 13 & \textless 1 \\
 & max. & 12 & 13 & 19 & \textless 1 \\
 \midrule
\multirow{2}{*}{\textbf{Whisk}} & avg. & \begin{tabular}[c]{@{}r@{}}261\\ 10 (cooldown)\end{tabular} & \begin{tabular}[c]{@{}r@{}}264\\ 10 (cooldown)\end{tabular} & \begin{tabular}[c]{@{}r@{}}263\\ 10 (cooldown)\end{tabular} & \begin{tabular}[c]{@{}r@{}}\textless 1\\ 171 (shuffle end)\end{tabular} \\
 & max. & \begin{tabular}[c]{@{}r@{}}263\\ 10 (cooldown)\end{tabular} & \begin{tabular}[c]{@{}r@{}}266\\ 10 (cooldown)\end{tabular} & \begin{tabular}[c]{@{}r@{}}265\\ 10 (cooldown)\end{tabular} & \begin{tabular}[c]{@{}r@{}}\textless 1\\ 185 (shuffle end)\end{tabular} \\
 \midrule
\multirow{2}{*}{\textbf{\begin{tabular}[c]{@{}c@{}}H. Sortition\end{tabular}}} & avg. & 277 375 & 513 750 & 2 013 006 & 441 \\
 & max. & 278 594 & 518 697 & 2 100 439 & 442 \\
\bottomrule
\end{tabular}
\end{table}

\subsection{Cost}

To evaluate the practical feasibility of the mechanisms, measurements were taken for block processing time, epoch processing time, and block size, shown in \autoref{tab:cost}. 
The results show that Whisk is approx. 20 to 30 times slower than the base selection due to cryptographic operations and periodic epoch shuffling; however, the overall duration remains generally negligible. 
In contrast, homomorphic sortition exhibited a dramatic increase in processing time, with costs rising steeply alongside the number of validators even with parallelization, making it unsuitable for Ethereum’s scale. 
While Whisk introduces moderate overhead, the extreme computational demands of homomorphic sortition render it impractical with current technology. 
In terms of network latency, block transmission remains negligible for Whisk and the base mechanism, but increases significantly for homomorphic sortition due to the large block size, especially in possible mechanism modifications selecting multiple proposers in advance, and the need for extended information sharing among participants required by the threshold FHE process.

\section{Discussion and Future Work}\label{sec:future-work}
Our findings underscore that while protection mechanisms increase the cost and complexity of an attack, they do not guarantee immunity. 
In particular, mechanisms that rely on a known candidate set are inherently vulnerable to attackers with sufficient foresight and resources. 
Additionally, client vulnerabilities, such as flaws in Ethereum execution clients~\cite{GethVuln}, may reduce the effort needed to perform advanced DoS attacks, bypassing the requirement of IP-identity mapping.

In summary, while Whisk and similar SSLE mechanisms show clear security benefits, they must be treated as a layer of deterrence rather than a foolproof defence. 
Their presence makes long-term attacks more expensive, less practical, and easier to detect over time. 
Moreover, by hiding the identity of proposers until the slot, these mechanisms significantly hinder or outright prevent MEV-based attacks, which typically rely on advance targeting and precise timing--conditions that are difficult to meet without executing a targeted DoS.
However, more robust protection is still needed in scenarios where attackers can invest long-term resources and remain covert.

While both mechanisms improve proposer anonymity, they differ significantly in computational cost. Whisk introduces moderate overhead (approx. 20–30 times the status quo) yet remains feasible within Ethereum’s timing constraints. In contrast, homomorphic sortition incurs extreme processing delays and scalability issues, making it impractical for real-world deployment without substantial optimization. However, efficiency could be improved by electing multiple leaders in advance using the Secret Leader Permutation (SLP) variant of the protocol, which could reduce the need to perform a full SSLE round for every slot. A high-level overview of these mechanisms and their properties is shown in~\autoref{tab:protection_comparison}.

\paragraph{\textbf{Future Work.}}
Future research should focus on attempting to mitigate the vulnerability of the known candidate list in Whisk that makes Ethereum even more prone to large-scale DoS attacks.
Integrating network-layer anonymization or adaptive shuffling strategies may help improve unpredictability. 
Additionally, refining the performance of homomorphic sortition--potentially via circuit optimizations or hybrid encryption schemes--could enable its real-world adoption. 
Given that some attacks exploit client-level vulnerabilities, future work should also investigate validator client hardening and the incorporation of failover strategies at the software level. 
Future work should also explore adapting the protocol for Ethereum by using the SLP variant to preselect multiple leaders once and investigating the use of committees to improve the scalability of threshold FHE operations.
Lastly, expanding the simulation framework to include more complex adversarial behavior and a network layer view would provide a more holistic view of protection mechanisms in adversarial environments.

\begin{table}[t]
\centering
\scriptsize
\caption{Comparison of Proposer Protection Mechanisms}\label{tab:protection_comparison}
\begin{tabularx}{\linewidth}{@{}lXXXX@{}}
\toprule
\textbf{Mechanism} & \textbf{DoS Protection} & \textbf{Computation Cost} & \textbf{Real-Time}\newline\textbf{Feasibility} & \textbf{Cons} \\
\midrule
\textbf{Whisk} & Moderate–High & Low & High & Known small\newline candidate set \\
\textbf{Homomorphic}\\ \textbf{Sortition} & Very High & Very High & Low & Scalability, latency \\
\bottomrule
\end{tabularx}
\end{table}

\section{Related Work}\label{relatedwork}
We organize the related work into four categories, described below.
\paragraph{\textbf{DoS Attacks \& Privacy}.}
Previous studies have speculated on the feasibility and implications of validator deanonymization and targeted attacks in Ethereum’s PoS consensus. 
Bürgel et al.~\cite{HOPRsniping} highlighted the realistic threat of validator sniping, demonstrating that IP addresses can be linked to public keys with high accuracy, and that minimal downtime is sufficient for a successor validator to extract MEV rewards. 
Similarly, the "Packetology" analysis~\cite{packetology} provided empirical evidence that network-layer data leaks enable associating validator indices with specific peer identities using timing patterns and gossip propagation. 
These works emphasized the potential for targeted DoS attacks and stressed the need for privacy-preserving mechanisms.

\paragraph{\textbf{SSLE Approaches.}}
Whisk~\cite{Whisk} introduces a shuffle-based SSLE mechanism for Ethereum that conceals proposers until block creation.  
It preserves strong ano\-ny\-mi\-ty despite partial node compromise and mitigates RANDAO~\cite{randao} biasing attacks.  
Whisk counters selling attacks, where validators prove future roles to MEV searchers, without adding complexity. 
Its bandwidth and latency overhead are modest, enabling feasible deployment.  
The authors compare Whisk with Dandelion++~\cite{dandelion} and earlier SSLE mechanisms, noting these often suffer from excessive overhead or weak MEV resistance.  
Whisk is tailored to Ethereum’s architecture, achieving high deployability with minimal coordination.  
Their evaluation positions Whisk as a leading SSLE candidate.
Buterin’s Simplified SSLE mechanism~\cite{swapornot} uses repeated size-2 blind-and-swap operations for low-overhead proposer privacy.  
It avoids complex cryptography and effectively obscures proposer identities under ideal conditions.  
Khovratovich~\cite{swapornotAnalysis} shows the scheme’s anonymity degrades in adversarial settings, where swaps can be traced or disrupted.  
Compared to Whisk, it is more vulnerable and requires more rounds to match privacy.
The Ouroboros~\cite{provable} family, underlying Cardano~\cite{cardano}, offers SSLE via private cryptographic lotteries.  
Each validator privately checks slot eligibility based on stake and public randomness.  
This prevents targeted attacks without public leader announcements, using cryptographic sortition instead of shuffling.

Homomorphic sortition~\cite{hSortition} enables encrypted leader selection to enhance proposer privacy without early identity disclosure. Though promising in theory, it remains unanalyzed or unadapted for Ethereum. Its impact on Ethereum’s fork-choice rule, latency, and network design is still unexplored.
LAKSA~\cite{reijsbergen2020laksa} targets secure, high-throughput PoS via stochastic leader election and committee confirmation.  
Inspired by Algorand~\cite{algorand},  DFINITY, and Randhound~\cite{hanke2018dfinity,syta2017scalable}, it enhances lightweight voting.  
Randomly sampled committees periodically vote on chain views~\cite{reijsbergen2020laksa}, improving robustness and scalability.

\paragraph{\textbf{SnSLE Approaches.}}
Secret Non-Single Leader Election (SnSLE) protocols, such as Algorand~\cite{algorand}, introduce inherent proposer privacy by allowing multiple potential proposers to be selected for a given slot, through the use of Verifiable Random Functions (VRFs)~\cite{VRF}.
However, this approach requires an additional round to determine the best block proposer from eligible candidates, which introduces additional communication overhead with linear complexity.
Vitalik Buterin’s SnSLE proposal~\cite{SnSLE} explores an alternative to SSLE based on probabilistic multi-proposer selection, inspired in part by Algorand~\cite{algorand}. 
In this approach, each validator independently learns if she is eligible to propose in a given slot, offering proposer privacy without requiring shuffling or encryption. 
While conceptually simpler, the design brings significant practical trade-offs. 
As noted by Whisk authors~\cite{Whisk}, non-single leader election complicates fork-choice rules, makes the protocol vulnerable to MEV time-buying attacks, and significantly increases networking overhead, especially in a sharded environment. 
Algorand addresses these challenges with protocol-level optimizations like smart gossip and timeouts but adapting such solutions to Ethereum would be nontrivial.

\paragraph{\textbf{Network-Level Anonymization.}}
A proposal to enhance validator anonymity using Dandelion++ and RLN~\cite{dandelion} aimed to obfuscate the origin of consensus messages by routing them through a private sub-network. 
While conceptually promising, the final analysis concluded that the approach is not feasible for Ethereum’s consensus layer due to strict latency constraints and high implementation complexity, especially under the tightened timing rules. 
The trade-offs ultimately limited its applicability despite its strong anonymity model. 

Polkadot’s Sassafras~\cite{sassafras}  combines an SSLE-based leader election with network-layer anonymity. It complements consensus-layer protection with mechanisms to conceal proposer network traffic, offering a more comprehensive defense against deanonymization and targeted attacks.
This strategy of providing privacy at the network level while selecting a single leader at the consensus layer is also utilized by other protocols, such as LAKSA~\cite{reijsbergen2020laksa}.

\paragraph{\textbf{Comparison with Our Work.}}
While prior work~\cite{Whisk,swapornotAnalysis,swapornot} has provided brief analyses and comparisons of individual proposer privacy mechanisms, there has not yet been a unified framework capable of evaluating multiple approaches under adversarial conditions. Our work addresses this gap by introducing a simplified Ethereum consensus model that enables systematic testing and comparison of SSLE mechanisms in hostile environments.

\section{Conclusion}\label{conclusion}
This work introduced a modular simulation framework for evaluating proposer privacy mechanisms in Ethereum under adversarial conditions. 
Using a simplified yet representative consensus model, we analyzed two SSLE mechanisms, Whisk and homomorphic sortition, against targeted and coordinated attacks. 
While both mechanisms show resilience to simple DoS attacks, neither withstands coordinated adversarial strategies and each presents practical limitations. 
Whisk introduces a vulnerability in which a subset of validators, known as the candidate set, is publicly exposed, making advanced attacks easier. 
In contrast, homomorphic sortition operates over the entire validator set, providing stronger security, but the computational requirements are currently too demanding for practical deployment. 
These findings offer a clearer understanding of the limitations and trade-offs in current SSLE designs and support a more informed development of proposer privacy mechanisms for Ethereum.
The framework’s extensibility enables future testing of additional designs, supporting continued development of effective and deployable proposer anonymity solutions for Ethereum.

\newpage

\bibliographystyle{splncs04}
\bibliography{bibliography}
\appendix
\section{Framework Implementation Details}
\label{appendix:implementation}

\begin{figure}[!b]
    	\centering
    	\vspace{-0.3cm}
        \includegraphics[width=1\textwidth]{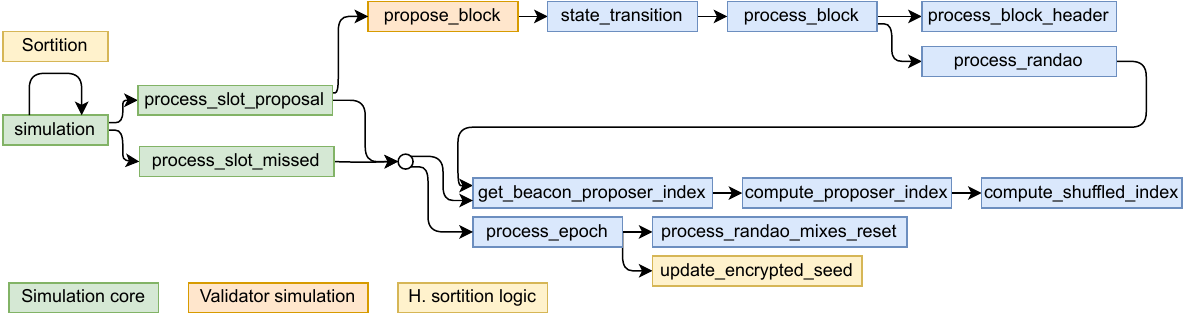}
    	\caption{The function call hierarchy in the consensus simulation as per the current consensus specification \cite{consensusSpec}.}
    	\label{fig:callsBase}
    \end{figure}

        \begin{figure}[!b]
    	\centering
        \vspace{-0.3cm}
    	\includegraphics[width=0.5\textwidth]{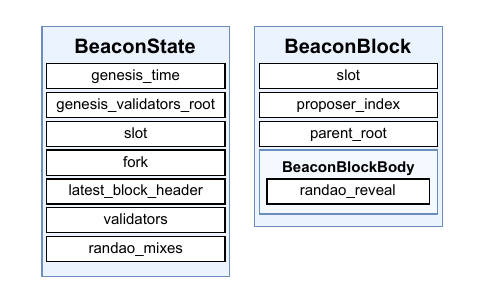}
        \vspace{-0.3cm}
    	\caption{The structure of the simplified beacon state and beacon block as per the current consensus specification \cite{consensusSpec}.}
    	\label{fig:beaconBase}
        \vspace{-0.5cm}
    \end{figure}

\subsection{Status Quo}
\autoref{fig:callsBase} describes the implementation of the simulation framework design as per the current consensus specification~\cite{consensusSpec} with the structure of the beacon state and the beacon block shown in~\autoref{fig:beaconBase}.

\subsection{Whisk}
\begin{figure}[h!]
    	\centering
    	\includegraphics[width=1\textwidth]{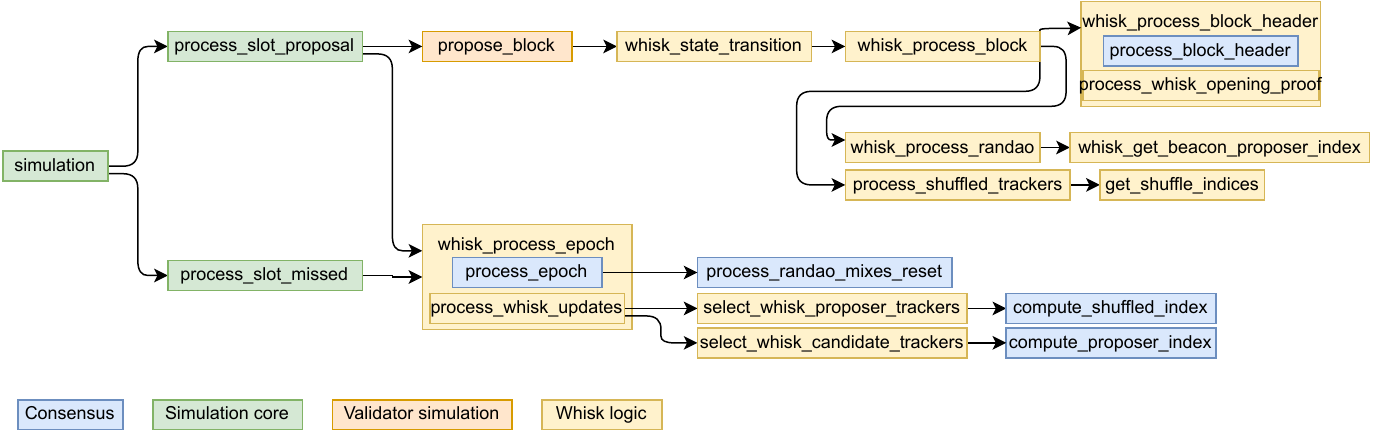}
    	\caption{The function call hierarchy in the consensus simulation as per the Whisk specification \cite{eip7441}.}
    	\label{fig:callsWhisk}
        \vspace{-0.8cm}
    \end{figure}

\begin{figure}[h!]
    	\centering
    	\includegraphics[width=0.5\textwidth]{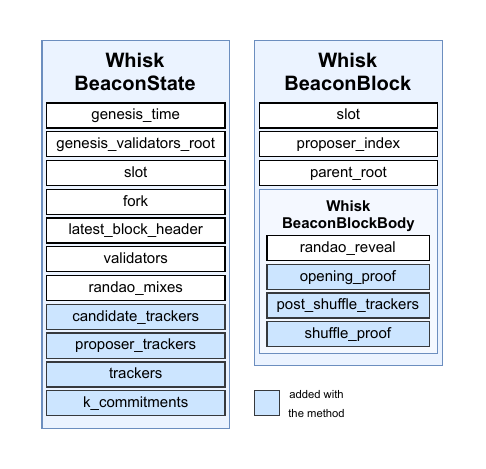}
        \vspace{-0.4cm}
    	\caption{The structure of the simplified beacon state and beacon block as per the Whisk specification \cite{eip7441}.}
    	\label{fig:beaconWhisk}
    \end{figure}

\autoref{fig:callsWhisk} and \autoref{fig:beaconWhisk} show the changed framework, beacon state and beacon block structure in Whisk implementation as per the Whisk specification draft~\cite{eip7441}.

    \subsection{Homomorphic Sortition}
\begin{figure}[h!]
        \centering
        \includegraphics[width=1\textwidth]{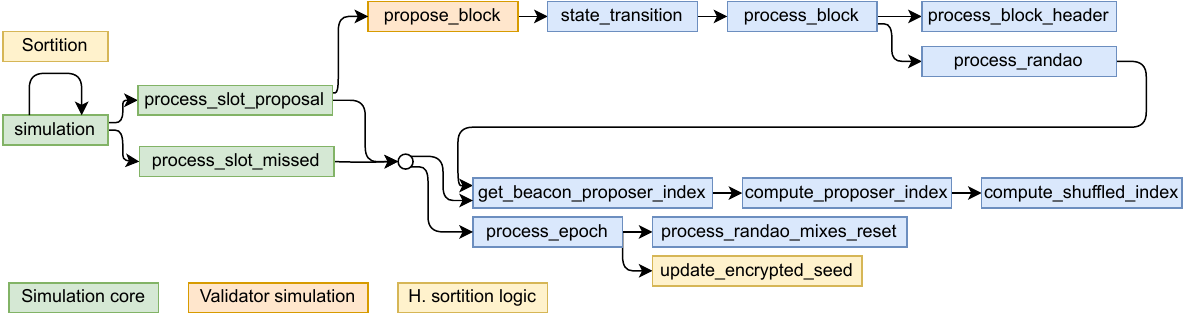}
    	\caption{The function call hierarchy in the consensus simulation using h. sortition.}
    	\label{fig:callsHS}
    \end{figure}

\begin{figure}[h!]
    	\centering
    	\includegraphics[width=0.5\textwidth]{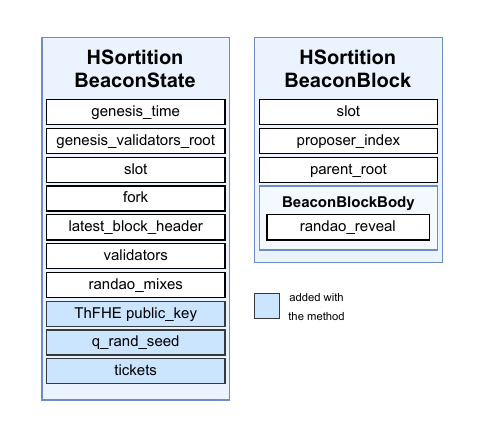}
        \vspace{-0.3cm}
    	\caption{The structure of the simplified beacon state and beacon block based on the own proposed homomorphic sortition design for Ethereum.}
    	\label{fig:beaconHSortition}
    \end{figure}

    \autoref{fig:callsHS} and \autoref{fig:beaconHSortition} show the changed framework, beacon state and beacon block structure in homomorphic sortition implementation, as designed for the simplified Ethereum consensus based on the article by Freitas et al.~\cite{hSortition}.

\end{document}